\documentclass[showpacs,preprint,aps]{revtex4}
\usepackage{graphicx}
\usepackage{amssymb}
\usepackage{amsmath}
\usepackage{mathrsfs}

\begin{document}
\setcounter{page}{1}
\title
{Three Photon Entanglement from Ortho-Positronium Revisited
\let\thefootnote\relax\footnote{Presented at 2nd Jagiellonian Symposium 
on Fundamental and  Applied Subatomic Physics, 
Krak\'ow, 2017 }
}
\author
{M. Nowakowski, D. Bedoya Fierro} 
\affiliation{Departamento de Fisica, 
Universidad de los Andes\\
Cra. 1E, 18A-10, Bogot\'a, Colombia}
\begin{abstract}
Entanglement of three photons from the
decay of ortho-positronium is re-analyzed. We use the full three body phase space
to write down the entangled states classified according to 
the spin directions of the ortho-positronium. Even in the case
when the spin is perpendicular to the decay plane, we find non-negligible
phases entering the entangled state. This has not been noticed before. We advocate a fixed quantization axis of the spin
for the sake of generality. A brief discussion of a three dimensional
formalism for photons, including correlations, versus a two dimensional one is given. 

\end{abstract}
\maketitle

\section{Introduction}
It is not often that a device designed principally for medical applications
turns out to be useful also in probing fundamental issues of physics. J-PET
(Jagellonian Positron Emission Tomograph) is a tomography machine based on positron-electron
annihilation with novel technological assesoire. It can be used to test CP/T and CPT violation
in the purely leptonic sector of positronium \cite{P1} and probably also in probing non-locality aspects
of quantum mechanics by studying the three photon entanglement from the decay of ortho-positronium (which
is a spin-1 bound state of electron-positron) \cite{P1, P2}.
It is the latter topic which we will discuss in this paper. Of course, many tests of locality
have been already performed which confirm the non-local nature of quantum mechanics.
That still other tests are beeing suggested (among other using the J-PET) lies in the
underlying positivistic philosophy of any natural theory which cannot be verified to one hundred per cent,
but falsified
by only one experimental result.
Coming back to the topic of three photon entanglement from ortho-positronium we will follow the
seminal paper \cite{ALP} which paved the ground for the subject. In doing so we
recover most of the results in \cite{ALP}, but find also important differences.
One of our motivation to revisit the subject is the interplay between relativistic dynamics/kinematics
and quantum entanglement.
\section{Kinematics and decay dynamics}
The spin-1 ortho-positronium decays into three photons 
where every photon is characterized by its momentum and polarization, 
$
\gamma_i=\gamma[\mbox{\boldmath $\epsilon$}(\hat{\mathbf{k}}_i, \lambda), \mathbf{k}_i]
$.
For circular polarization we have the polarization vectors
\begin{eqnarray} \label{polariz}
  \mbox{\boldmath $\epsilon$}(\hat{\mathbf{k}}_i, \lambda)&=&-\frac{\lambda}{\sqrt{2}}\biggl(\cos\theta_i \cos\Phi_i-i
  \lambda\sin\Phi_i, 
\cos\theta_i\sin\Phi_i
\nonumber \\
&+&i\lambda\cos\Phi_i, -\sin\theta_i\biggr) 
\end{eqnarray}
with $\lambda= \pm 1$ and the angles define the direction of each momentum by
$
\hat{\mathbf{k}}_i=\left(\cos\Phi_i\sin\theta_i, 
\sin\Phi_i\sin\theta_i, \cos\theta_i \right)
$
The energy-momentum conservation at rest reads
$
\mathbf{k}_1+\mathbf{k}_2 +\mathbf{k}_3=0
$
and
$
k_1 +k_2+k_3=m\simeq 2m_e
$
The momentum conservation defines a plane in which the photons move. The plane changes its orientation from event to event.
The unit vector
$
\hat{\mathbf{n}}=\frac{\mathbf{k}_1 \times \mathbf{k}_2}{|\mathbf{k}_1 \times \mathbf{k}_2|}
$
is perpendicular to the plane. Often one chooses
$
\hat{\mathbf{n}}=\hat{\mathbf{z}}
$ and the question arises if we can do it for one event or
for the whole sample.

Briefly, the matrix element ${\cal M}$ is
$
{\cal M}=-\sqrt{2}V_3
$
for $S_z=0$ and
$
{\cal M}=\pm V_1+iV_2
$
for $S_z=\pm 1$. The vector function $\mathbf{V}$ 
is a lengthy expression \cite{Itzy, ALP}, but an important one for the entanglement. We therefore  
give it in full length here
\begin{eqnarray} \label{matrix3}
&&\mathbf{V}(\mathbf{k}_1, \lambda_1; \mathbf{k}_2, \lambda_2; \mathbf{k}_3, \lambda_3)=
\nonumber \\
&&(\lambda_1-\lambda_2)(\lambda_2+\lambda_3)\mbox{\boldmath $\epsilon$}^*(\hat{\mathbf{k}}_1, \lambda_1)\left[
\mbox{\boldmath $\epsilon$}^*(\hat{\mathbf{k}}_2, \lambda_2)\cdot \mbox{\boldmath $\epsilon$}^*(\hat{\mathbf{k}}_3, \lambda_3)\right]
\nonumber \\
&+&(\lambda_2-\lambda_3)(\lambda_3+\lambda_1)\mbox{\boldmath $\epsilon$}^*(\hat{\mathbf{k}}_2, \lambda_2)\left[
\mbox{\boldmath $\epsilon$}^*(\hat{\mathbf{k}}_3, \lambda_3)\cdot \mbox{\boldmath $\epsilon$}^*(\hat{\mathbf{k}}_1, \lambda_1)\right]
\nonumber \\
&+& (\lambda_3-\lambda_1)(\lambda_1+\lambda_2)\mbox{\boldmath $\epsilon$}^*(\hat{\mathbf{k}}_3, \lambda_3)\left[
\mbox{\boldmath $\epsilon$}^*(\hat{\mathbf{k}}_1, \lambda_1)\cdot \mbox{\boldmath $\epsilon$}^*(\hat{\mathbf{k}}_2, \lambda_2)\right]
\nonumber
\end{eqnarray}
This function encodes the whole dynamics including
the entanglement. For instance,
$
V(\mathbf{k}_1, \pm; \mathbf{k}_2, \pm; \mathbf{k}_3, \pm)=0
$ 
\section{The three photon entanglement}
On the other hand,
we find for instance
$
\mathbf{V}(\mathbf{k}_1, +; \mathbf{k}_2, +; \mathbf{k}_3, -)
=2\mbox{\boldmath $\epsilon$}^*(\hat{\mathbf{k}}_3, -) f_{12}
$
with $f_{ij}=\left[1-\hat{\mathbf{k}}_i\cdot \hat{\mathbf{k}}_j \right]$. 
This leads directly to the coefficients of the entangled states classified according
to the spin projections.
For 
$S_z=0$ and ${\cal M}=-\sqrt{2}V_3$
we obtain
$\epsilon_3^*(\hat{\mathbf{k}}_i, \lambda=\pm)=-\sin\theta_i$. Therefore, 
in this case the unnormalized three parties entangled state comes out as
\begin{eqnarray} \label{e1}
|\Psi\rangle_{S_z=0} &=&  \gamma_0 \left[|++-\rangle -|--+\rangle \right] 
+  \beta_0 \left[|+-+\rangle -|-+-\rangle \right] \nonumber \\
&+&  \alpha_0 \left[|-++\rangle -|+--\rangle \right]  
\end{eqnarray}
with the coefficients 
\begin{equation} \label{e1c}
\gamma_0= \sin \theta_3 f_{12}, \,\, 
\beta_0= \sin \theta_2  f_{13}, \,\,
\alpha_0=\ \sin \theta_1 f_{23}. 
\end{equation}
On the other hand for
$S_z=\pm 1$ and  ${\cal M}=\pm V_1+iV_2$ we have
\begin{equation} \label{a1}
\epsilon_1^*(\hat{\mathbf{k}}_i, \lambda=\pm) +i\epsilon_2^*(\hat{\mathbf{k}}_i, \lambda=\pm)= e^{i\Phi_i}
(\cos \theta_i \pm 1)
\end{equation}
This leads to the  entangled state of the form
\begin{eqnarray} \label{e2}
|\Psi\rangle_{S_z=\pm 1}&=&  \gamma_{\pm}^{(1)} |++-\rangle +  \gamma_{\pm}^{(2)}|--+\rangle 
+  \beta_{\pm}^{(1)} |+-+\rangle + \beta_{\pm}^{(2)}|-+-\rangle \nonumber \\  
&+&  \alpha_{\pm}^{(1)} |-++\rangle + \alpha_{\pm}^{(2)}|+--\rangle   
\end{eqnarray}

with 
\begin{eqnarray}
\gamma_{\pm}^{(1)}&=&  e^{\pm i\Phi_3} \left(\cos  \theta_3 -1\right) f_{12},\,\, 
\gamma_{\pm}^{(2)}=  e^{\pm i\Phi_3} \left(-\cos  \theta_3 -1\right) f_{12} 
\nonumber \\
\beta_{\pm}^{(1)}&=&    e^{\pm i\Phi_2} \left(\cos  \theta_2 -1\right) f_{13}, \,\, 
\beta_{\pm}^{(2)} =    e^{\pm i\Phi_2} \left(-\cos  \theta_2 -1\right) f_{13} \nonumber \\
\alpha_{\pm}^{(1)}&=&    e^{\pm i\Phi_1} \left(\cos  \theta_2 -1\right) f_{23},\,\,
\alpha_{\pm}^{(2)}=    e^{\pm i\Phi_1} \left(-\cos  \theta_2 -1\right) f_{23}. 
\nonumber 
\end{eqnarray}
For the first case ($S_z=0$), it is only if we  choose a coordinate system such that  $\hat{\mathbf{n}}=\hat{\mathbf{z}}$
($\sin \theta_i=1$) that
the expression becomes simpler and coincides with the one given in \cite{ALP} for $S_z=\pm 1$. 
In the case of $S_z=\pm 1$ we do not
recover the expression in \cite{ALP} for $S_z=0$. Even if we choose the spin
quantization axis in the $\hat{\mathbf{n}}$ direction, we differ by the phases  $e^{\pm i\Phi_i}$.
Apart from that, we have a different assignation for  $S_z=0, \pm 1$. 
This means that our coefficients depend explicitly on the coordinates ($\Phi_i$) 
even if we take $\theta_i=\pi/2$. Dependence on $\Phi_i$ has also been noticed in \cite{P2}
which appeared after the talk was given.  

How important are the phases? If, in addition to choosing the z-axis perpendicular to the
three photon plane, we make a rotation of the x-y coordinates, we can get rid of one phase.
Factorizing a second phase (which becomes global) we are certainly left with one relative phase.
One can demonstrate already in a simpler system (two particle entanglement) that a phase will play a role
in correlation functions which enter the Bell's inequalities. Deforming a spin- singlet by writing
\begin{equation} \label{deformed}
|00\rangle_{\alpha}=\frac{1}{\sqrt{2}}\left(|\uparrow \rangle |\downarrow \rangle - e^{i\alpha} |\downarrow \rangle |\uparrow \rangle
\right)
\end{equation}
the correlation function becomes
\begin{eqnarray} \label{deformed2}
&&_{\alpha}\langle 00|(\hat{\mathbf{a}}\cdot \mathbf{S}^{(1)}\hat{\mathbf{b}}\cdot \mathbf{S}^{(2)})|00 \rangle_{\alpha}\nonumber = \nonumber \\ 
&&-\frac{1}{4} \left[\cos \alpha \cos \theta + a_zb_z(1-\cos \alpha )- (\mathbf{a} \times \mathbf{b})_z \sin \alpha \right]
\nonumber
\end{eqnarray}
with the internal variable $\cos \theta =\hat{\mathbf{a}} \cdot \hat{\mathbf{b}}$
and (external) variable $a_i$ and $b_i$  (coordinates of the unit vectors).
We can then safely state that in correlation functions the phases  $e^{\pm i\Phi_i}$ which we encountered 
will certainly play a role.
Incidentally, with linearly polarized photons reference \cite{Kwiat}
writes a two photon entanglement as
$
\frac{1}{\sqrt{2}}\left[ |H\rangle |V \rangle + e^{i\alpha}|V \rangle |H \rangle \right]
$
where $|H \rangle =(1,0)$  and $|V \rangle =(0, 1)$
and the matrices used in the correlations are the Pauli matrices. 
A related  question comes then into mind. Shall we choose $\hat{\mathbf{n}}=\hat{\mathbf{z}}$?
For one single event this is certainly possible. If we do this for every subsequent event we will not be able
to classify the entanglements according to the spin projections of the positronium 
$S_z=0, \pm 1$ since we keep on changing the quantization axis then. If we group together, say 
$|\Psi \rangle_{S_z=0}$ with $|\Psi \rangle_{S_{z'}=0}$ 
we might be comparing and classifying  wrongly. 
The condition that the state factorizes (giving essentially rise to two-particle entanglement) is
given by
\begin{equation} \label{fact}    
\beta=0,\,\, \gamma=\pm \alpha;\,\,
\gamma=0,\,\, \beta=\pm \alpha ;\,\,
\alpha=0, \,\, \gamma=\pm \beta. 
\end{equation}
For $\theta_i=\pi/2$ choosing one of the three cases above we get
a configuration at the edge of the allowed phase space: 
two collinear momenta and with third one anti-parallel, e.g.
for $\gamma=0$ and $\beta=\alpha$. 
For an arbitrary spin quantization axis for the spin-1 ortho-positronium there are other configurations
of momenta for which the state factorizes.

In a three-party entanglement there are two
non-bisseparable classes: W-class and GHZ-class (Greenberger-Horne-Zeilinger)  
which cannot be transformed into each other
by local operations \cite{Vidal}. Generically, one writes
\begin{equation} \label{GHZ}
|GHZ \rangle =\frac{1}{\sqrt{2}}\left[ |000 \rangle + |111 \rangle \right],\,\,
|W \rangle = \frac{1}{3}\left[ |001 \rangle + |010 \rangle + |100 \rangle \right] 
\end{equation}
If we have a three-party entanglement $|\phi \rangle =\sum_{ijk}c_{ijk}|ijk \rangle $
an invariant measure of the entanglement is the so-called
hyper-determinant   
\begin{equation} \label{hdet}
0\le Hdet(c_{ijk})=c_{000}^2c_{111}^2+....+ ...c_{000}c_{110}c_{001} +...\le 1/4
\end{equation}
If $Hdet=0$ and the state does not factorize then we have
a W-class entanglement. If we choose $\hat{\mathbf{n}}=\hat{\mathbf{z}}$
the condition  $Hdet=0$ leads to factorization, i.e.,
the configuration of two collinear and one anti-parallel three momentum which we had before.
Since by our choice of the z-axis we might lose some generality, could it be
that we can reach the W-class in accordance with the energy-momentum conservation?  
The constraint of the energy-momentum conservation makes the problem
more complex than anticipated.  We just outline the first few steps.
First we perform a change of basis, from circular to linear
polarization $|\pm \rangle =|R/L \rangle$, $|H/V \rangle =|0/1 \rangle$ with
$
|\pm \rangle =\frac{1}{\sqrt{2}}\left[ |0 \rangle +i|1 \rangle \right]
$
Then, for instance
\begin{eqnarray} \label{change}
|\Psi \rangle _{S_z=0}&=& \left(\alpha_0 +\beta_0-\gamma_0\right) |010 \rangle
+\left(\alpha_0 -\beta_0-\gamma_0\right) |100 \rangle \nonumber \\
&+&\left(-\alpha_0 -\beta_0+\gamma_0\right) |001 
+\left(\alpha_0 +\beta_0+\gamma_0\right) |111 \rangle \nonumber 
\end{eqnarray}
It is easy to calculate the hyperdetereminant in this basis. The result is
\begin{equation}
Hdet=\left(-\alpha_0 +\beta_0 -\gamma_0 \right) \left(\alpha_0 -\beta_0 -\gamma_0 \right)\nonumber \\ 
\cdot \left(-\alpha_0 -\beta_0 +\gamma_0 \right) \left(\alpha_0 +\beta_0 +\gamma_0 \right) \nonumber 
\end{equation}
The main question regarding $Hdet=0$ is  whether this is possible in accordance with energy-momentum conservation (see also \cite{P2}).
This remains to be our future task.
\section{The two dimensional versus the three dimensional formalism}
What do we mean exactly when we write, say $|++- \rangle$ ? Obviously, it is the tensor product
$|++- \rangle = |+\rangle \otimes |+ \rangle \otimes - |\rangle$
But what exactly is, say $|+ \rangle $ or what is its representation
in a finite dimensional Hilbert space? Consider the following:
1. We have a two-level system given by the two degrees of freedom of the photon polarization 
$\lambda =\pm 1$
and
2. The photons carry momenta and are entangled in their polarizations which depend on the directions of the momenta, hence
$
|+ \rangle =|+ \rangle_1 =|\hat{\mathbf{k}}_1, + \rangle=|\mbox{\boldmath $\epsilon$}(\hat{\mathbf{k}}_1, +1) \rangle
$
But the polarization vectors are three dimensional objects. 
The mismatch between the number of qubits $\pm$ and the dimension of the state
vector comes from the fact that photons are vector particles, but strictly speaking,  do not have spin (defined only in the rest frame).
The standard correlation functions are of the form
\begin{equation}
\langle  \Psi|(\hat{\mathbf{a}}\cdot \mbox{\boldmath $\sigma$}^{(1)})
(\hat{\mathbf{b}}\cdot \mbox{\boldmath $\sigma$}^{(2)})(\hat{\mathbf{c}}\cdot \mbox{\boldmath $\sigma$}^{(3)})|\Psi \rangle
\end{equation}
with $\sigma_i$ the two dimensional Pauli matrices
(this correlation enters inequalities like the Mermin or Svetlichny inequality).
This implies that the state vectors are also two dimensional. 
This goes back to the so-called Jones formalism where all polarization states
are represented as two dimensional objects. In general
$
|\psi \rangle =(\cos \phi e^{i \alpha_x}, \sin \phi e^{i\alpha_y})
$  
If the difference between the phases is  
$\pi/2$ and $\cos \phi = \sin \phi$
we get the circular polarization states
$
\frac{1}{\sqrt{2}}(1, i), \,\,\, \frac{1}{\sqrt{2}}(1, -i)
$
This is in accordance with the polarization vectors 
if $\theta =0$ (and not 
$\pi/2$ ) i.e. we choose $\hat{\mathbf{k}}=
\hat{\mathbf{z}}$. Indeed, we have then
\begin{equation} \label{pol1}
\mbox{\boldmath $\epsilon$}(\hat{\mathbf{k}}, +)=\frac{e^{-i\Phi}}{\sqrt{2}}
(1, i, 0) ,\,\,
\mbox{\boldmath $\epsilon$}(\hat{\mathbf{k}}, -)=\frac{e^{-i\Phi}}{\sqrt{2}}
(1, -i, 0)
\end{equation}
which is effectively two dimensional. However, we cannot choose
for all photons $\hat{\mathbf{k}}=
\hat{\mathbf{z}}$ especially when we have chosen once 
$\theta_i=\pi/2$.
For two photons with $\mathbf{k}_1 + 
\mathbf{k}_2=0$, say in the decay of para-positronium with the entanglement
$
|\Psi \rangle_{para}=\frac{1}{\sqrt{2}}\left[|++ \rangle -|-- \rangle \right]
$\  
we can do that if we choose $\hat{\mathbf{k}}_i=
\hat{\mathbf{z}}$. 
With three photons we will have to choose a more appropriate
formalism and  operators (corresponding to, say circular polarizations).
We can do that by going to the adjoint representation of 
$SU(2)$
i.e., 
$(S_i)_{jk}=-i\epsilon_{ijk}$
The eigenvectors to $S_3$
are
$
\frac{1}{\sqrt{2}}(1, \pm i, 0)
$
which are the circular polarizations states in a plane
perpendicular to the z-axis. 
A third eigenvector $(0, 0, 1)$ is possible,
but the photon will not have it.
The eigenvectors to 
$S_1$ and 
$S_2$ are
$
\frac{1}{\sqrt{2}}(0, \pm i, 1), \,\, \frac{1}{\sqrt{2}}(\pm i,0, 1) 
$
respectively. The first state describes circular polarized
photons in a plane perpendicular to the x-axis etc. 
It makes then sense to consider $\hat{\mathbf{a}} \cdot \mathbf{S}$
and to calculate expectation values of the form
\begin{equation}
\langle  \Psi|(\hat{\mathbf{a}}\cdot \mathbf{S}^{(1)})
(\hat{\mathbf{b}}\cdot \mathbf{S}^{(2)})
(\hat{\mathbf{c}}\cdot \mathbf{S}^{(3)})|\Psi \rangle
\end{equation}
\section{Conclusions}
The physics of the three photon entanglement from positronium has all
the physical ingredients which makes the twentieth and the beginning of the twenty first century physics:
the positron necessary to form the positronium is a decay product of a nuclear
decay, it forms a leptonic bound state whiose physics is well
understood in the framework of Quantum-Electro-Dynamics and finally
the three body entanglement has been a subject of intense studies in the last two decades.
In this contribution we have re-analyzed the emtangled state of three photons
from ortho-postronium in full generality and found some differences as compared to \cite{ALP}.\\
{\bf Acknowledement}:
We thank A. Botero and C. Viviescas for useful discussions and comments.


\end{document}